\documentclass[a4paper, 9pt, twocolumn]{article}
\usepackage[utf8]{inputenc}
\usepackage[T1]{fontenc}
\usepackage{graphicx}
\usepackage{longtable}
\usepackage{wrapfig}
\usepackage{rotating}
\usepackage[normalem]{ulem}
\usepackage{amsmath}
\usepackage{amssymb}
\usepackage{capt-of}
\usepackage{hyperref}
\newcommand{\fullcitenourl}[2][]{%
\AtNextCitekey{%
\clearfield{url}%
\clearfield{urlyear}%
\clearfield{date}%
\clearfield{urlday}%
\clearfield{urlmonth}%
\clearfield{note}%
\clearfield{eprint}%
\clearfield{doi}%
\clearfield{isbn}%
\clearfield{issn}%
\clearfield{publisher}%
\clearfield{location}%
\clearfield{series}%
\clearfield{volume}%
\clearfield{number}%
\clearfield{pages}%
}%
\ifblank{#1}{%
\citeauthor{#2}. \normalfont{\citetitle{#2}} in \citeyear{#2}.%
}{%
\citeauthor{#2}. \normalfont{\citetitle{#2}} in  \citeyear{#2}, p. #1.%
}%
}

\usepackage{xr}

\RequirePackage[normalem]{ulem} 
\RequirePackage{color}\definecolor{RED}{rgb}{1,0,0}\definecolor{BLUE}{rgb}{0,0,1} 

\usepackage{microtype} 

\usepackage{amsmath,amsfonts,amsthm} 

\PassOptionsToPackage{svgnames}{xcolor}
\usepackage{awesomebox}
\usepackage{cleveref}

\usepackage[size=small, labelfont=bf, up, textfont=it]{caption} 

\usepackage{booktabs} 

\usepackage{lastpage} 

\usepackage{graphicx} 

\usepackage{enumitem} 
\setlist{noitemsep} 

\usepackage{sectsty} 
\allsectionsfont{\usefont{OT1}{phv}{b}{n}} 
\usepackage{cuted}

\usepackage{geometry} 

\usepackage{wrapfig}

\usepackage[T1]{fontenc} 
\usepackage[utf8]{inputenc} 

\usepackage{XCharter} 

\usepackage{fancyhdr} 
\fancypagestyle{firstpage}{ 
	\fancyhf{}
}

\usepackage{titling} 

\newcommand{\HorRule}{\color{DarkGoldenrod}\rule{\linewidth}{1pt}} 

\pretitle{
	\vspace{-50pt} 
  \fontsize{14}{16}\usefont{OT1}{phv}{b}{n}\selectfont 
	\color{DarkRed}\raggedright  
}

\posttitle{
  \par\vskip 8pt
} 

\preauthor{} 

\postauthor{ 
  \par\HorRule
  \vspace{-35pt}
}

\usepackage{lettrine} 
\usepackage{fix-cm}	

\newcommand{\initial}[1]{ 
	\lettrine[lines=3,findent=4pt,nindent=0pt]{
		\color{DarkGoldenrod}
		{#1}
	}{}%
}
\usepackage{xstring} 

\newcommand{\lettrineabstract}[1]{
	\StrLeft{#1}{1}[\firstletter] 
	\initial{\firstletter}\textbf{\StrGobbleLeft{#1}{1}} 
}

\makeatletter
\newcommand{\PaperTitle}[1]{\def\@PaperTitle{#1}}
\newcommand{\Archive}[1]{\def\@Archive{#1}}
\newcommand{\Authors}[1]{\def\@Authors{#1}}
\newcommand{\JournalInfo}[1]{\def\@JournalInfo{#1}}
\newcommand{\Abstract}[1]{\def\@Abstract{#1}}
\newcommand{\Keywords}[1]{\def\@Keywords{#1}}
\makeatother
\usepackage{titling,kantlipsum}

\usepackage[backend=bibtex,style=authoryear,natbib=true, style=numeric]{biblatex} 

\usepackage[autostyle=true]{csquotes} 

\DeclareSourcemap{
  \maps{
    \map{
      \step[fieldsource=url,
            match=\regexp{/books\.google\.},
            fieldset=url, null]
    }
  }
}

\usepackage[english]{babel}
\usepackage{authblk}

\usepackage{cuted}
\usepackage{changepage}
\usepackage{cleveref}
\usepackage{breqn}
\usepackage{chngcntr}
\usepackage{xr}
\usepackage{appendix}
\usepackage{svg}
\definecolor{mygold}{HTML}{f1c232}
\definecolor{mygrey}{HTML}{595959}
\definecolor{mygrey2}{HTML}{3E3E3E}
\definecolor{myred}{HTML}{b92e3f}

\definecolor{mypurple}{HTML}{A799B7}
\definecolor{myteal}{HTML}{93B7BE}

\definecolor{myterra}{HTML}{3d405b}
\definecolor{mygreen}{HTML}{72a190}
\definecolor{mygray2}{HTML}{595959}
\definecolor{mygray3}{HTML}{495057}

\usepackage{tikzpagenodes}
\definecolorseries{chcolor}{rgb}{grad}[rgb]{.95,.85,.55}{3,11,17}
\resetcolorseries{chcolor}

\usepackage{cleveref}

\usepackage[most]{tcolorbox}
\newtcolorbox[
auto counter,
number freestyle={\noexpand\arabic{\tcbcounter}~\noexpand\mytitle},
crefname={box}{boxes}]%
{infobox}[2][]{
colbacktitle=mygreen,
colback=white!80!gray!,
coltitle=white,
colframe=mygreen,
boxrule=0.5pt,
fonttitle=\bfseries,
boxed title style={colframe=mygreen},
code={\def\mytitle{#2}},
title=Box \thetcbcounter,%
breakable=false,
sidebyside=false,
top=0pt,
bottom=0pt,
#1,
}

\newtcbtheorem[auto counter]{deff}{Definition}{
lower separated=false,
colback=white!80!gray,
colframe=white,
fonttitle=\bfseries,
colbacktitle=mygreen,
coltitle=white,
enhanced,
boxed title style={colframe=mygold},
attach boxed title to top left={xshift=0.5cm,yshift=-2mm},
}{def}

\newtcbtheorem[auto counter]{mymyth}{Myth}{
lower separated=false,
colback=white!80!gray,
colframe=white,
fonttitle=\bfseries,
colbacktitle=mygreen,
coltitle=white,
enhanced,
boxed title style={colframe=mygold},
attach boxed title to top left={xshift=-.25cm,yshift=-2mm},
}{myth}

\newtcbtheorem[auto counter]{mymyth2}{Myth}{
lower separated=false,
colback=white!80!gray,
colframe=white,
fonttitle=\bfseries,
colbacktitle=mygreen,
coltitle=white,
enhanced,
boxed title style={colframe=mygold},
attach boxed title to top left={xshift=-.25cm,yshift=-2mm},
}{myth}
\usepackage{multicol}

\usepackage{cuted}
\author{Casper van Elteren}
\date{}
\title{Three Myths in Complexity Science -- and How to Resolve Them}

\begin{document}
\maketitle \noindent
\begin{strip}
\vspace{-3em}
\lettrineabstract{Complex systems permeate our world, from ant colonies to financial markets, captivating scientists with their emergent behaviors. Yet, the field of complexity science is often shrouded in myths and misunderstandings. This paper challenges three pervasive myths: that the whole is more than the sum of its parts, that complex systems defy reductionist approaches, and that complexity requires many constituents. By critically examining these beliefs, we provide a modern interpretation of complex systems, redefining emergence and its relation to system behavior. We argue that emergence stems not from magical ingredients but from constraints on degrees of freedom, producing outcomes different from—not greater than—the sum of parts. This perspective reconciles reductionist and holistic approaches, offering a more nuanced understanding of complexity. Our analysis provides a compass for navigating the intricate waters of complexity science, bridging theoretical foundations with practical applications and paving the way for more rigorous and insightful research in this captivating field.
}
\end{strip}

Complex systems  are  fascinating and  ubiquitous
phenomena  that  exist  all  around   us  –  think  of  ants
collaborating   to  forage   for  food,   starlings  forming
mesmerizing murmurations  at dusk,  or the dynamic  flows of
traffic  on   our  city  streets.  These   classic  examples
illustrate the  delicate web  of interactions  that underpin
complex systems,  where individual components  work together
in harmony to  produce emergent properties that  are said to
produce something greater  than the sum of  their parts. The
beauty  and  complexity  of these  systems  have  captivated
scientists  and philosophers  for  centuries, inspiring  new
approaches to  understanding and addressing  pressing global
issues.

\begin{figure}
\centering
\includegraphics[width=\linewidth]{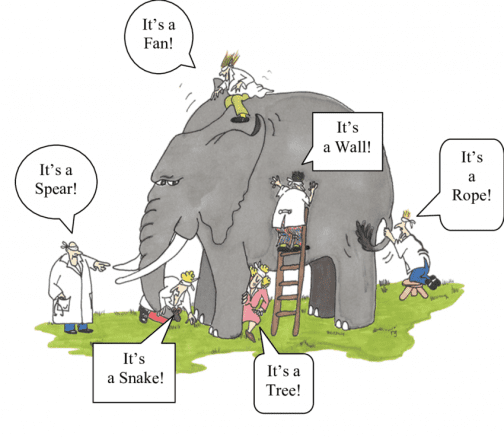}
\caption[Caption for LOF]{\label{fig:blind-men}Complex systems can be compared to the legend of the blind men and the elephant, where each man perceives only a part of the elephant, leading to incomplete and varied understandings. This cartoon illustrates that understanding complex systems requires integrating different perspectives to form a complete picture.\footnotemark}
\end{figure}

However, despite the apparent intuitive grasp we may have of
these  systems,  the  scientific  understanding  of  complex
systems         is          far         from         settled
\cite{Ladyman2013,Estrada2023,Hadorn2008,Karaca2022,Thurner2016}.
While examples like ants  foraging, starlings murmuring, and
traffic  flowing may  evoke  a sense  of understanding,  the
concept  of  complex systems  remains  debated  and lacks  a
universally   accepted   definition   (as   illustrated   in
\cref{fig:blind-men}).

Definitions of complex  systems typically involve properties
ill-defined properties, such as the requirement for ``many''
elements for complexity to arise, the need for memory or the
emergence of robust  order \cite{Phelan2001}. Such inquiries
immediately raise  questions, casting doubt on  the validity
of the field. The goal of  this work is not to get entangled
in semantics  or attributing complexity to  the observations
of an external  observer (\cref{fig:blind-men}). Rather, the
idea  is to  give  a modern  interpretation  of how  complex
systems are understood.

To provide  clarity for the  field of complexity  science, I
aim to resolve three myths that permeate the field:

\begin{mymyth}{}{}
\raggedright The whole is more than the sum of its parts.
\end{mymyth}

\begin{mymyth}{}{}
\raggedright The behavior of complex systems cannot be understood through reductionistic approaches.
\end{mymyth}

\begin{mymyth}{}{}
\raggedright Complexity requires the interaction of many constituents.
\end{mymyth}

To  discuss these  myths,  I will  first  examine how  parts
relate  to  the  whole  within  the  context  of  complexity
science. We will dig deep  and provide historical context to
the field  and how the myths  came to be. I  will argue that
the   common   belief   that   complexity   science   defies
reductionism  and  produces   something  ``more''  leads  to
logical   inconsistencies   and   a  reliance   on   magical
ingredients to  explain macroscopic structures.  This belief
is  often  conflated  with  the  concept  of  emergence.  By
providing  concise,  operational   definitions  of  emergent
properties  and  emergent  behavior,  I  will  clarify  this
misconception.  With these  definitions, we  will understand
that complex systems do not offer ontological novelty in the
strong sense  but rather  impose restrictions on  degrees of
freedom, producing outcomes that  are different from the sum
of their parts.

\section{Path Dependency in the Science of Complexity}
The field of complexity science  has its roots in scientific
history, dating  back to the  establishment of the  Santa Fe
Institute  in  1984   \cite{HistorySantaFe}.  This  landmark
institution  brought together  experts from  various fields,
including physics, biology, economics, and computer science,
to integrate  and consolidate knowledge on  complex systems.
The Santa Fe Institute played  a pivotal role in shaping our
understanding of  these intricate phenomena by  fostering an
interdisciplinary approach that encouraged collaboration and
innovation   among  researchers.   Through  its   work,  the
institute laid  the foundation  for new theories  and models
that  would eventually  be applied  to real-world  problems,
from economic forecasting to climate modeling.

The early days of the field of complexity science, concerned
itself with so-called "wicked problems" – problems that defy
reductionistic analysis and emphasize  the importance of the
web    of    relations   underpinning    complex    problems
\cite{Rittel1973}.   The  field   aimed  to   be  in-between
different  approached,  integrating  upon existing  work  to
produce  novel approaches  to  understand so-called  complex
systems. This required a fundamental shift in thinking, from
viewing  complex systems  as mere  aggregates of  individual
parts to  recognizing them  as emergent entities  that arise
from the interactions and relationships between those parts.
By embracing  this new  perspective, complexity  science was
able to tackle problems  that had previously been considered
unsolvable or at least extremely difficult.

\begin{figure}[htbp]
\centering
\includegraphics[width=.9\linewidth]{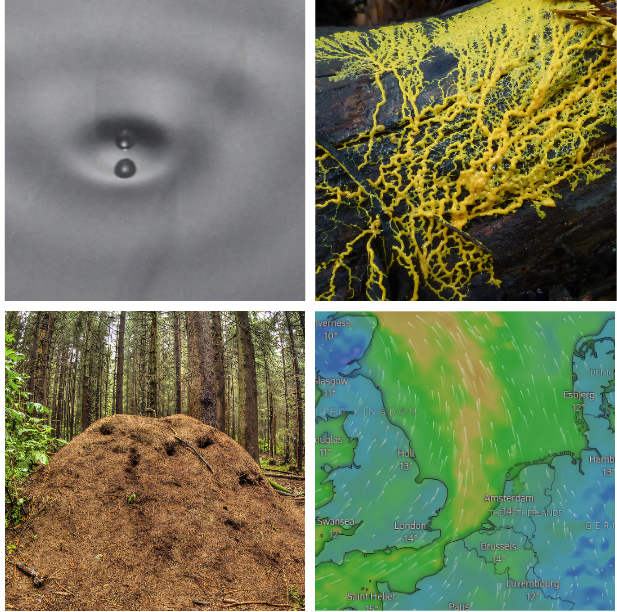}
\caption{\label{fig:typical-complex-systems}Complex systems exist at multiple scales, ranging from the microscopic interactions of elementary particles (experiment oil droplets in a double slit experiment emulating a ``pilot wave'', top left) to the intricate inner workings of cells (slime mold, top right), the dynamics of human societies, and the vast interdependencies within ecosystems (anthill, bottom left) and climate (wind flows, bottom right). This illustration highlights how these systems, despite their diverse scales, share common characteristics such as interconnectivity, emergent behavior, and adaptability, illustrating the universality and complexity of interactions across different levels of organization.}
\end{figure}

The term complex system was  used to differentiate it form a
simple system.  In essence, a  system is defined as  a tuple
\(S   =  (P,   R)\)   comprising  a   collection  \(P\)   of
interconnected  parts, related  through a  mapping \(R(P)\).
The  parts  themselves  can   represent  various  levels  of
description relevant  to understanding,  potentially leading
to  a  nested  hierarchy   of  systems.  Alternatively,  the
elements might  capture the fundamental, atomistic  level of
description necessary to comprehend natural phenomena.

The   word   ``complex''   comes   from   the   Latin   word
\emph{complexus},  meaning ``embraced''  or ``encompassed''.
This,   in  turn,   is   derived   from  ``com-''   (meaning
``together'') and  \emph{plectere} (meaning ``to  weave'' or
``to braid'')  \cite{dictComplex}. Taken together  a complex
system etymological means a  collection of interwoven parts.
Colloquialy,     ``complex''     carries    two     distinct
interpretations.

Firstly,  it   denotes  something   that  is   difficult  to
understand.  For example,  sending  a rocket  into space  or
providing a comprensive model  to accurately predict whether
climate  predictions  are  distinctly  more  difficult  than
computing the square root of, say, 49. A complex problem may
require  highly specialized  skill  or  knowledge to  solve,
consist  of  many different  parts  making  it difficult  to
address than simpler problems.

\footnotetext{Blind men and the elephant cartoon by G. Renee Guzlas, artist.}

Secondly, it  refers the  interconnectedness of  many parts.
This second interpretation may refer to being able to access
all the information necessary to solve the problem. That is,
the problem seems  to defy reducing it to  a simpler problem
with less parts and less information necessary to solve. The
parts are an  integral part of the system,  and the observed
system behavior  does not emerge from  straightforward rules
governing individual components.

Both  interpretations permeate  the  science of  complexity,
which is  defined by the  interrelatedness of parts  and the
inherent difficulty in predicting  the behaviors that emerge
from their interactions \cite{Nicolis2012}.

The literature  can thus be  summarized by stating  that the
term  \emph{complex systems}  is commonly  used to  describe
systems  consisting  of  parts   that  produce  behavior  or
properties   greater   than   the  sum   of   their   parts.
Understanding or predicting the behavior of a complex system
requires analyzing how  the interaction structure (\emph{its
  form})  relates  to  its  behavior  (\emph{its  function})
\cite{Huneman2012,Anderson1972}.   These  interactions   can
involve   simple   pairwise  connections   or   higher-order
relationships,   which  may   be   weighted  or   temporally
dependent, adding layers of complexity.

The  key point  is  that the  interrelatedness  of a  system
produces  phenomena  distinct  from the  properties  of  its
individual parts  studied in isolation. Complex  systems are
closely  tied  to  the   concept  of  emergence,  where  new
behaviors and properties arise  from the interactions within
the system.  The relationship  between components  cannot be
reduced to  mere linear relations, as  linearity would imply
superposition, which could be  understood by analyzing parts
separately. Instead, the study of complex systems requires a
holistic  approach to  grasp the  emergent behavior  arising
from the intricate web of interactions.

\section{How Do the Parts Relate to the Whole?}
\label{sec:orgba10741}
What makes a system complex is  not just the number of parts
it has,  but rather  the web  of interactions  between those
parts.  It's the  way these  individual components  interact
with each other that gives  rise to behaviors that fascinate
many. With the enormous  success of classical mechanics, the
reductionistic  way  of thinking  led  many  to believe  for
centuries    that   reducing    the   thing-to-be-understood
(\emph{explandum})  to  a  part description  was  sufficient.  By
decomposing   behavior  in   terms   of  the   sum  of   the
interactions, we  could grow  closer to  understanding \emph{how}
the explandum arises.

According  to some  complexity  scientists, complex  systems
inhabit a special niche  that challenges this reductionistic
way of thinking. As G. H. Lewes articulated in 1879:

\vspace{1em}
\begin{quote}
\emph{Every resultant is either a sum or a difference of the co-operant forces; their sum, when their directions are the same – their difference, when their directions are contrary. Further, every resultant is clearly traceable in its components, because these are homogeneous and commensurable. It is otherwise with emergents, when, instead of adding measurable motion to measurable motion, or things of one kind to other individuals of their kind, there is a co-operation of things of unlike kinds. The emergent is unlike its components insofar as these are incommensurable, and it cannot be reduced to their sum or their difference.}
\vspace{-0.5em}
\begin{flushright}
\fullcitenourl[368]{Lewes1879}
\end{flushright}
\end{quote}
\vspace{1em}

According to Lewes, interrelated  parts can create something
beyond  mere aggregation.  For example,  surface tension  of
water is not due to the mere aggregation of water molecules.
Rather, wetness  arises from the hydrogen  bonds between the
slightly polarized water molecules  that create the property
of water tension. Likewise, it  is not the grand ensemble of
neuron  activity   that  produces  consciousness,   but  the
interconnected coordination of billions of neurons.

This  sense  of wonder  and  enchantment  is not  unique  to
Lewes's quote,  but is a  common reaction when  people first
encounter  complex  systems. It  \emph{feels}  as  if these  intricate
networks  of  interactions  are   performing  some  kind  of
``magical'' trick, where disparate components come together to
create  something entirely  new and  unexpected. It  further
complicates the question  of how parts relate  to the whole,
as   it   seems   to   challenge   reductionist   approaches
(\cref{infobox:reductionism}).

The  property by  which a  whole (system)  exhibits behavior
that is different  from the properties of the  parts is also
known as emergence, to which we turn next.

\section{Emergence in Complex Systems - Delineating Etymology, Property, and Behavior}
\label{sec:orgb3f8b84}
Emergence  is best  understood as  the idea  that the  whole
system  exhibits  properties  or   behaviors  that  are  not
directly    attributable    to    its    individual    parts
\cite{Bedau2008,Butterfield2011,deHaan2006}.  In other  words,
the system  transcends the  sum of its  components, creating
something novel  or something  that is difficult  to predict
from  the parts  alone.  Through  often simple  interactions
among its  parts, novel  and sometimes  unexpected behaviors
emerge at the system level.

Despite  the intuitive  appeal  of emergence,  it remains  a
concept  that is  often  viewed  as troublesome,  especially
within             philosophical             discourse \cite{Hulswit2005,Kim2006,Padgett2012}.

\begin{strip}
\begin{infobox}[label = infobox:reductionism]{Explanations in Complex Systems: Reductionism, and Holism}
Complexity   scientists   face    the   daunting   task   of
understanding  and  modeling  complex systems.  Two  primary
approaches  emerge:  reductionism and  holism.  Reductionism
involves breaking down these systems into their constituent
parts and analyzing each piece separately to explain a
high-dimensional phenomenon in terms of a lower-dimensional
object that is equally well described.

In  contrast, holistic  thinking considers  the system  as a
whole,  focusing  on  its emergent  properties  rather  than
individual   components.  While   both  approaches   aim  to
understand the  natural world, they differ  fundamentally in
the level of compression  that can be achieved. Reductionism
seeks to  compress complex phenomena  into lower-dimensional
objects, whereas  holism suggests that this  compression may
not  always  be  possible, requiring  consideration  of  the
entire system as a focal point.

However,  rather   than  being  mutually   exclusive,  these
approaches can be seen  as complementary strategies lying on
a continuum (\cref{fig:compression}). When  a complex system can
be understood through a singular, e.g. one dimensional,  description, it is fully reduced
to a part. As interactions become more significant, however,
detail   is  required   for  explanation.   In  the   limit,
considering the  whole system becomes necessary,  leading to
holism.

\vspace{1em}
\centering \includegraphics[width = \linewidth]{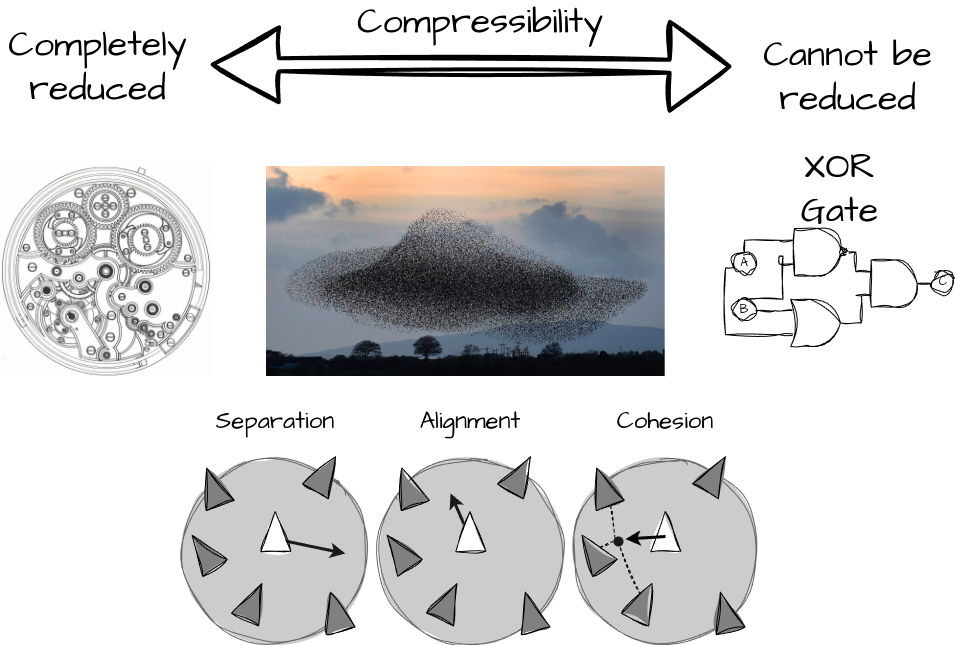}
\captionof{figure}{The tension between reductionism and holism can be resolved by considering both approaches in terms of the level of compression. The mechanics of a watch can be completely reduced to the summation of the workings of the parts (image credit: Andrea Muratore). As the complexity increases, more complex patterns such a flocking could be reduced to simple rules combined with nearest neighbor interactions. Murmurations, for example, can be understood through three simple rules: separation, alignment, and cohesion. Separation ensures individuals avoid crowding their neighbors, alignment means they steer towards the average heading of nearby flockmates, and cohesion keeps them close to the center of the group. At maximum incompressibility, we are left with a system in which the whole is necessary such as parity checks or XOR gates.}
\label{fig:compression}
\vspace{1em}
\end{infobox}
\end{strip}

The  concept   of  emergence   has  its  roots   in  ancient
philosophy, notably dating back to Aristotle \cite{Cohen2021}.
However,  it  wasn't until  the  late  19th and  early  20th
centuries that  the term  ``emergence'' became widely  used to
describe the relationship between  constituent parts and the
whole.  The   term  itself  derives  from   the  Latin  verb
\emph{emergere}, meaning ``to rise out''  or ``to come forth,'' which
is formed from the prefix  \emph{ex-} meaning ``out,'' and the verb
\emph{mergere},  meaning  ``to  dip''  or  ``to  sink.''  Originally,
emergence referred to the act  of rising out or coming forth
from a submerged or hidden state. In modern usage, emergence
generally  denotes  the  process   of  becoming  visible  or
apparent, such as a body emerging from water.

Aristotle's  philosophy  provides  an  early  framework  for
understanding emergence, particularly through his concept of
hylomorphism, which posits that  everything is a compound of
matter (\emph{hyle})  and form  (\emph{morphe}). He believed  that the
form of an  object is not just its shape  but its essence or
what it is meant to be.  The form gives matter its structure
and purpose,  suggesting an emergent relationship  where the
whole  is more  than just  a  collection of  its parts.  For
example, a  house is not  a mere collection of  sticks; they
are cut  and constructed with  great diligence to  carry the
weight of a roof, provide doors to walk through, and so on.

Emergence, however,  comes in different forms,  notably weak
and strong  emergence. Some  might argue that  these systems
show weak emergence, where the emergent properties can still
be  traced  back  to  the  interactions  of  the  individual
components   \cite{Chalmers2002,Rupe2023,Rosas2024}.  However,
for  systems  that  exhibit strong  emergence,  there  would
indeed not  be such  a mapping  back from  the whole  to the
parts. To date, no such system has ever been seen in nature.
Consciousness is  often considered to show  strong emergence
\cite{McLaughlin2023},  however, this  view  is questioned  by
some and  is not not settled  \cite{Chalmers1995a}. Still, the
notion  of different  forms of  emergence is  useful to  get
closer to an understanding of what is meant by emergence.

Building on Aristotle's foundational ideas, Henri Poincaré's
contributions  in the  late  19th and  early 20th  centuries
brought  renewed  attention  to the  concept  of  emergence,
emphasizing    its    relevance   in    modern    scientific
thought \cite{Heinzmann2024}. Poincaré asserts that:

\vspace{1em}
\begin{quote}
\emph{The whole is not merely the sum of its parts, but it is something more than this.}
\vspace{-0.5em}
\begin{flushright}
\fullcitenourl{Poincare1905}
\end{flushright}
\end{quote}
\vspace{1em}

His  work  highlighted  the  importance  of  looking  beyond
reductionist approaches, which attempt to understand complex
systems  solely by  dissecting their  individual components.
Instead, he advocated for  a holistic view, recognizing that
interactions between parts can  lead to novel and unexpected
outcomes.  This  idea  has been  profoundly  influential  in
various scientific  disciplines, such as systems  theory and
complexity  science, where  researchers  explore how  simple
rules  and interactions  give rise  to complex  and emergent
behaviors.  In essence,  Poincaré's insights  bridge ancient
philosophical    ideas    with    contemporary    scientific
understanding,   reinforcing   the  notion   that   emergent
properties  of  a  system  contribute to  its  identity  and
function in ways that transcend  the mere aggregation of its
parts.  His  work continues  to  inspire  and inform  modern
theories  of  complexity   and  systems,  demonstrating  the
enduring   relevance  of   the  concept   of  emergence   in
understanding the natural world.

\subsection{Novelty of What?}
\label{sec:org73cbfc1}
The problem with  Poincaré's addition of the  word ``more'' is
that we are left with a problem: what is more to the system?
In the  context of Poincaré's  work, we can deduce  that the
``more''  refers to  the  patterns and  system behaviors  that
emerge     through     non-linear    dynamics,     producing
difficult-to-predict properties based on initial conditions.
Prior to Poincaré,  the success of reductionism  led many to
believe that  given enough measurement prowess,  we could—in
principle—deduce the future trajectories  of many systems by
applying the laws of physics.

In the  concept of  emergence, however,  it is  important to
note  that Aristotle  never intended  to invoke  a sense  of
``more.'' In fact, his texts always refer to the property that
the whole  is \emph{different} from  the parts (see the  quote at
the beginning of this book).  To state something is more, we
need to explicate what this novelty is and how it arises.

The problem with novelty in the context of emergence lies in
its  ambiguous nature.  Defining what  constitutes ``novelty''
can  be  challenging,  as   it  implies  that  the  emergent
properties  are not  merely a  sum or  rearrangement of  the
existing    parts,   but    something   fundamentally    new
\cite{Garud2015}. This raises questions  about the origins and
mechanisms of these new properties. For instance, how does a
complex  system give  rise to  behaviors or  characteristics
that were not present in the individual components? That is,
is it impossible to understand  the surface tension of water
in terms of the electromagnetic forces within the individual
water  molecule?  For  the known  observable  systems,  this
simply is not the case.

Moreover, compared to the degrees of freedom that individual
components  possess, complex  systems  are actually  limited
systems \cite{Odor2004,Thompson2004}.   This  might  seem
counterintuitive,  but mutual  constraints  within a  system
generate  correlations  and  dependencies that  prevent  the
system  from  exploring its  full  degrees  of freedom.  For
example,  in a  biological  organism,  individual cells  are
constrained by  the need  to function coherently  within the
larger system of the body. These constraints limit the range
of possible  behaviors and  interactions, leading to  a more
organized and structured system.

Novelty,  in  this  sense,  is not  necessarily  a  gain  of
function but rather a lack of it (\cref{infobox:novelty}). The
emergent   properties  arise   from   the  limitations   and
constraints imposed on the  components, which restrict their
behavior in specific ways.  This restricted behavior results
in new patterns and properties that we perceive as emergent.
Thus, the novelty  seen in complex systems  is a consequence
of  the interplay  between  parts and  the constraints  they
impose on  each other, rather  than an outright  creation of
new functions or capabilities.

\begin{strip}
\begin{infobox}[label = infobox:novelty]{Complex Systems and Emergent Novelty}

\begin{wrapfigure}{r}{0.5\textwidth}
\centering \includegraphics[width = 0.85\linewidth]{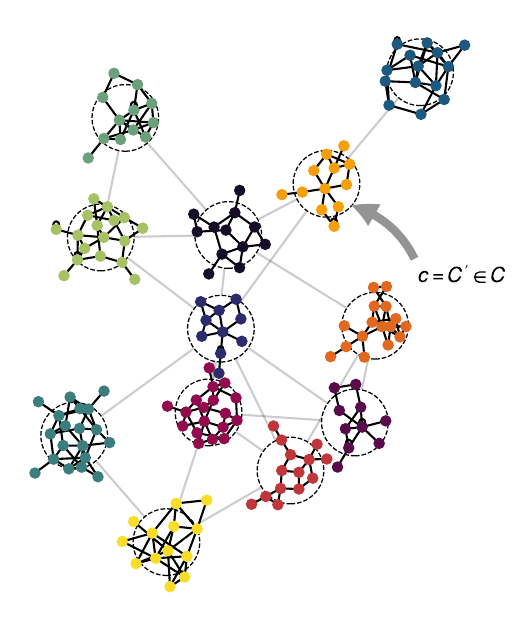}
\captionof{figure}{\label{fig:nested_hierarchy} Novelty can be understood of system $C$ by creating an augmented system $C^'$, effectively creating a nested hierarchy. The properties of each ``element" is produced through mutual constraints of the lower levels. Within the simpler system $C$, however, novelty is best understood as the diminishing degrees of freedom of the parts created through local correlations. }
\end{wrapfigure}
  Complex systems are often less complex than their parts, a
  counterintuitive  idea  that   creates  tension  with  the
  novelty  we experience  in everyday  life. Surely  novelty
  exists,  right?  Let's  consider  a complex  system  $C$  as
  depicted in \cref{fig:nested_hierarchy}.

  $C$ is  represented with  sufficient detail  to understand
  how the  system's behavior  emerges. Assume elements  of C
  can be  in one of $Q$  distinct states. The full  range of
  possible   states   is   then   described   by   $Q^{|C|}$
  configurations.  However,   many  complex   systems  don't
  explore this full exponential space. Instead, they inhabit
  a much smaller subspace,  producing correlations over time
  and/or space.

  We can  extend $C$ to a  finer-grained resolution $C^{'}$,
  which  itself becomes  a system  with its  own interaction
  structure,  statespace, and  potentially different  rules.
  Importantly, each element $c \in C$ now becomes a subsystem,
  created by correlations within the subspace $(Q')^{|c|}$.

  By repeating this procedure,  we can understand novelty as
  a causal influence  propagating through these hierarchies.
  ``Novelty''    (restrictions)    emerges   from    mutual
  constraints cascading \emph{all the way down} these nested
  levels  of  organization.  Naturally for  other  kinds  of
  dynamic  we imagine  this process  extend over  continuous
  space.

\end{infobox}
\end{strip}

\subsection{How to Understand Emergence? From Property to Behavior}
\label{sec:org43a4537}
Complex  systems exhibit  emergent behavior,  but what  does
that mean?  To understand  emergent behavior, we  must first
identify when  a complex  of elements possesses  an emergent
property. This  requires a fundamental understanding  of how
to approach complex systems.

There  are  (at least)  three  distinct  ways to  understand
complex systems \cite{Bertalanffy1950,Bertalanffy1969}. First,
we can view them as mere collections of individual elements,
where the whole is simply the sum of its parts. For example,
a  complex made  up  of  three poodles  is  understood as  a
complex  of   poodles  \cite{Georgiou2003}.  Second,   we  can
categorize  them  based  on  the  characteristics  of  their
constituent elements. For instance,  a complex consisting of
one poodle, one Dalmatian, and one Alsatian is understood as
a  complex of  dogs because  it includes  elements from  the
species  ``dogs''.  However,   these  approaches  neglect  the
crucial aspect of relationships between elements.

A  more  comprehensive   understanding  of  complex  systems
emerges  when  we  consider the  relationships  between  the
elements.  For example,  a  complex made  up  of a  palomino
stallion, his mare, and their pony is understood as a family
of palomino  horses. In this  case, it is necessary  to know
the relationships  between the  stallion, mare, and  pony to
define  the complex  as a  family of  palomino horses.  This
relational   understanding   recognizes  that   a   system's
properties arise not just  from individual elements but from
the dynamic  interactions between them. This  perspective is
essential for analyzing complex  systems, as it acknowledges
that the  system's behavior  and emergent  properties emerge
from the interactions between its components.

The  level  of  explanation  needed  to  understand  complex
systems  varies  significantly  based  on  context  and  the
individual's expertise \cite{Marr1982}. Take, for example, the
phenomenon  of opinion  polarization. At  a basic  level, we
recognize  it as  the creation  of divides.  However, deeper
analysis  prompts us  to ask  how exactly  this polarization
occurs. This requires delving  into specific mechanisms that
drive   polarization,  such   as   societal  structures   or
historical contexts.

To  truly  grasp the  behavior  and  emergent properties  of
complex    systems,   a    comprehensive   explanation    is
essential—one  that   considers  the   dynamic  interactions
between elements,  as discussed earlier. The  sufficiency of
the explanation  regarding these  mechanisms depends  on the
specific  questions asked  and the  goals of  understanding.
Recognizing  the need  for different  levels of  explanation
allows us  to tailor our responses  effectively, providing a
more  accurate and  comprehensive  understanding of  complex
systems.

From this, we can define what an emergent property is and how it relates to behavior in complex systems.

\begin{deff}{Emergent property}{}\label{def:em
}
A characteristic of a system that exists by virtue of the interrelation of the parts. Addition or removal of the parts or interrelation will change the property (or make it disappear).
\end{deff}

\begin{deff}{Emergent behavior}{} \label{def:em
}
Emergent behavior is  an emergent property expressed as a function of time. Specifically, emergent behavior is the temporal (and potentially transient) behavior that arises from the interactions of a complex system's components.
\end{deff}

Armed with these definitions, we can  now turn to the use of
complex systems in the field of complexity science.

\section{Emergence in Modern Complexity Science}
\label{sec:org0382ef4}
In complexity  science, emergence  is intrinsically  tied to
the  interactions  produced by  a  system  or a  complex  of
elements.  It  refers  to  systemic  regularities,  such  as
behavior or  structure, that arise from  the interactions of
individual   elements.   This  modern   perspective   echoes
Aristotle's  idea that  emergent properties  arise from  the
complex  organization  and  interaction of  parts  within  a
whole. We conclude by revisiting the myths discussed
at the beginning.

\textbf{Myth 1: The whole is more than the sum of its parts.}
The whole is  not merely more than the sum  of its parts; it
is different. The apparent novelty of emergent systems is an
illusion;  in reality,  they  exist due  to  a reduction  in
degrees  of freedom,  creating correlations  among elements.

\textbf{Myth 2: Emergent Systems Evade Reductionist Analysis.}
Furthermore, complex systems  demonstrating (weak) emergence
do  not evade  reductionist analysis.  Complexity scientists
provide  explanations by  relating  system  behavior to  the
interactions of its parts,  thus revealing the mechanisms by
which  the  emergent  properties arise.  This  understanding
allows  for  intervention-based  approaches to  examine  how
changes in parts  affect the whole system.

\textbf{Myth 3: Emergent Systems Requires Many Entities.}
Emergent systems can arise  in any system with interactions,
eliminating  the  need for  a  large  number of  interacting
entities  often   suggested  in  the   literature.  Emergent
properties  and  behaviors   can  exist  through  non-linear
interactions between as few as two entities.

In  modern complex  systems science,  the study  of emergent
properties  has profound  implications  for various  fields,
from  biology and  ecology to  sociology and  economics. For
instance,  the   emergence  of  consciousness   from  neural
interactions  in  the  brain, the  formation  of  ecosystems
through  species interactions,  and market  dynamics arising
from  individual economic  agents' behaviors  all illustrate
how complex systems science  seeks to understand and predict
emergent   phenomena.   Researchers  utilize   computational
models, network  theory, and  data analytics  to investigate
these systems,  aiming to uncover the  underlying principles
that govern emergent behavior.

We have therefore reached  a modern understanding of complex
systems, where a system is considered complex if it exhibits
an emergent  property, and  its dynamics result  in emergent
behavior.

In  conclusion,  I  have  sought to  identify  and  critique
persistent flaws in complexity science. Despite the vastness
of  the literature  and the  evolution of  perspectives over
time, I hope my contributions shed new light on this field ---
providing, if you  will, a rudder and a  compass to navigate
the complex waters  of this discipline. Just  as Leonardo da
Vinci cautioned  against practice without theory,  this work
aims  to  balance  theoretical  foundations  with  practical
applications in complexity science. Given the scope of these
discussions, it was impossible to cover all the discourse on
complex systems  that has occurred over  the past centuries.
Some  may notice  obvious omissions,  such as  computational
complexity,    statistical     mechanics,    self-organizing
criticality, and others. By focusing on certain myths, I aim
to advance our understanding of complex systems by examining
a common root issue: language  may shift the focus away from
what is truly important to understanding something separate.
While the  use of  language and  semantics is  important, it
should not  obstruct progress.  Consensus should  only arise
out of necessity \footnote{A similar discussion can be found
  in the  discourse about defining  what Japanese art  is in
  \emph{Killing Commendatore}, chapter  9 by Haruki Murakami
  (2017),  and the  definition  of a  game in  Philosophical
  Investigations by Ludwig Wittgenstein (1953).}. 

Looking ahead,  there is much  to be done to  further refine
the borders of complexity  science. The sketch provided here
implies  that  a  system   may  exhibit  multiple  forms  of
emergence  through   properties  and   different  behaviors.
Furthermore,  how  the results  and  parts  result in  these
properties  is not  explicated  other than  pointing to  the
interactions and  stating that these interactions  should be
more than linear mixing.

I owe  a great  intellectual debt to  the pioneers  who have
shaped the study of complexity. The framework presented here
sets the  stage for  future exploration of  how interactions
between components  affect the behavior and  form of complex
systems.


\printbibliography
\end{document}